\documentclass[aps,prd,nofootinbib,twocolumn,superscriptaddress,preprintnumbers]{revtex4}

\usepackage{amssymb}
\usepackage{amsmath}
\usepackage{epsfig}
\usepackage{hyperref}
\usepackage{breakurl}
\usepackage{bm}
\usepackage{color}
\usepackage{tikz}
\usetikzlibrary{decorations.markings}

\makeatletter
\def\simgt{\mathrel{\lower2.5pt\vbox{\lineskip=0pt\baselineskip=0pt
           \hbox{$>$}\hbox{$\sim$}}}}
\def\simlt{\mathrel{\lower2.5pt\vbox{\lineskip=0pt\baselineskip=0pt
           \hbox{$<$}\hbox{$\sim$}}}}
\makeatother

\newcommand{\be}{\begin{equation}}
\newcommand{\ee}{\end{equation}}
\newcommand{\eq}[2]{\be\begin{aligned}#1 \label{#2}\end{aligned}\ee}

\newcommand{\Ref}[1]{Ref.~\cite{#1}}

\newcommand{\Eq}[1]{Eq.~\eqref{#1}}
\newcommand{\Eqs}[2]{Eqs.~\eqref{#1} and \eqref{#2}}

\newcommand{\MEFT}{M_{\rm EFT}}

\newcommand{\II}{{\cal I}}

\def\topbotatom#1{\hbox{\hbox to 0pt{$#1\bot$\hss}$#1\top$}} \newcommand*{\topbot}{\mathrel{\mathchoice{\topbotatom\displaystyle} {\topbotatom\textstyle} {\topbotatom\scriptstyle} {\topbotatom\scriptscriptstyle}}}

\begin{document}

\title{From Scattering Amplitudes to Classical Potentials \\ \smallskip in the Post-Minkowskian Expansion}

\author{Clifford Cheung}
\affiliation{Walter Burke Institute for Theoretical Physics,
    California Institute of Technology, Pasadena, CA 91125}
\author{Ira Z. Rothstein}
\affiliation{Department of Physics, Carnegie Mellon University, Pittsburgh, PA 15213}
\author{ Mikhail P. Solon}
\affiliation{Walter Burke Institute for Theoretical Physics,
    California Institute of Technology, Pasadena, CA 91125}
    
\begin{abstract}

We combine tools from effective field theory and generalized unitarity to construct a map between on-shell scattering amplitudes and the classical potential for interacting spinless particles.   For general relativity, we obtain analytic expressions for the classical potential of a binary black hole system at second order in the gravitational constant and all orders in velocity.  Our results exactly match all known results up to fourth post-Newtonian order, and offer a simple check of future higher order calculations.   By design, these methods should extend to higher orders in perturbation theory.  

\end{abstract}

\preprint{\hbox{CALT-TH-2018-031}} 

\maketitle

\section{Introduction}

The theory of scattering amplitudes has revealed unique insights into the structure of quantum field theory (QFT) and inspired powerful new tools for calculation.   While phenomenological applications have largely centered on high-energy colliders, an effort has emerged to connect the amplitudes program to the physics of gravitational waves, which were recently discovered at LIGO~\cite{Abbott:2016blz}.    

Unfortunately, any attempt at bridging these subjects is immediately confounded by the fact that a binary black hole inspiral is quite dissimilar from black hole scattering.  The latter is a transient interaction of widely separated black holes which are effectively free before and after the event.  The former describes objects bound in quasi-circular orbit by a classical conservative potential, together with the dissipative radiation-reaction force induced by gravitational wave emission.  

There is a long history of mapping scattering observables to the classical gravitational potential, {\it e.g.}~see the seminal work of \cite{Iwasaki:1971vb,Hiida:1972xs} as well as more recent treatments \cite{Neill:2013wsa, Bjerrum-Bohr:2013bxa, Vaidya:2014kza, Damour:2016gwp, Cachazo:2017jef, Guevara:2017csg,Damour:2017zjx, Bjerrum-Bohr:2018xdl}.  In this paper we unify ideas from effective field theory (EFT) and generalized unitarity to systematize this procedure for a general QFT of spinless particles~\cite{Neill:2013wsa,Bjerrum-Bohr:2013bxa}. To begin, we construct an EFT for two non-relativistic (NR) scalars which interact via the classical potential $V$.  Since the two-particle on-shell amplitudes in the EFT and full theory are equal, {\it i.e.} $\MEFT = M$, we can determine $V$ order by order in perturbation theory.

Of course, on-shell methods like generalized unitarity vastly simplify amplitude calculations (see Refs.~\cite{Bern:2011qt,Elvang:2015rqa} and references therein).  In this approach, $M$ is expressed not in terms of Feynman diagrams but rather as a sum of scalar integrals weighted by scalar integral coefficients which are rational functions of the external momenta.  

Our main results are summarized in \Eq{eq:result_2PM}, which recasts the coefficients $c$ of the classical  potential in terms of the scalar integral coefficients $d$ in a general QFT at leading and next-to-leading order in the interaction strength. For general relativity (GR), we obtain  \Eqs{eq:result_gravity_2PM}{eq:Vpos}, which are new analytic expressions for the potential at second post-Minkowskian (2PM) order, {\it i.e.}~at ${\cal O}(G^2)$ and at {\it all} orders in velocity.    These equations are physically equivalent to all state-of-the-art results, which extend to fourth post-Newtonian (4PN) order~\cite{Damour:2014jta, Bernard:2015njp,Jaranowski:2015lha}. Since our results include information at all orders in the PN expansion, they may be useful for checking future higher order calculations.  
The present work goes beyond previous calculations of the 2PM amplitude~\cite{ Bjerrum-Bohr:2013bxa, Cachazo:2017jef, Bjerrum-Bohr:2018xdl} by deriving an explicit mapping to the 2PM potential.

This work introduces several new methods.  First, we show how calculations are drastically simplified when the classical limit is taken at the earliest possible stage of the computation. This is implemented by a simple power counting scheme in large angular momentum $J \gg 1$, together with a restriction on loop momenta to the so-called potential region of kinematics. Copious quantum mechanical contributions are thus truncated at the integrand level while complicated four-dimensional integrals are reduced to far simpler three-dimensional ones.

Second, we introduce the method of ``integrand subtraction'' to effectively eliminate three-dimensional integrals which can be quite complex due to infrared singularities.  In this approach, the {\it difference} of the integrands in the full theory and EFT are similar to those encountered in NR GR~\cite{Goldberger:2004jt,Gilmore:2008gq} and easily integrate to purely rational functions of the external kinematics.

Third, we show how gauge-dependent quantities like the classical potential can be compared by computing gauge-invariant on-shell scattering amplitudes without the need for constructing explicit coordinate transformations or wrangling with equations of motion ambiguities.

\section{Effective Field Theory}

\noindent {\bf Definition.}  An EFT for NR scalar fields $A$ and $B$ is described by the action
$S = \int dt \;( L_{\rm kin} +  L_{\rm int})$, where
\eq{
L_{\rm kin} =& \phantom{ {}+{} } \int_{\bm k}   \, A^\dagger(-\bm k) \left(i\partial_t - \sqrt{\bm k^2 + m_A^2}\right) A(\bm k)\\
& + \int_{\bm k}    \, B^\dagger(-\bm k) \left(i\partial_t - \sqrt{\bm k^2 + m_B^2}\right) B(\bm k) \, ,
}{eq:L_kin}
is the kinetic term and the interaction term is~\cite{Neill:2013wsa}
\eq{
L_{\rm int} =  -\int_{{\bm k},{\bm k'}}  \, V(\bm k ,\bm k') \, A^\dagger(\bm k') A(\bm k) B^\dagger(-\bm k') B(-\bm k) \, .
}{eq:L_int}
Here $\int_{{\bm k_1} \cdots {\bm k_n}} = \int {d^3 \bm k_1 \over (2\pi)^3} \cdots {d^3 \bm k_n \over (2\pi)^3}$ and the Feynman vertex $V(\bm k ,\bm k') $ is the potential in the center of mass frame.  

\medskip

\noindent {\bf Classical Limit}.  The above EFT is obtained from the full theory by integrating out massless force carriers mediating near-instantaneous interactions and taking the NR limit,  $|\bm k| , |\bm k'| \ll m_{A,B}$. By definition, these potential modes have energies parametrically less than their momenta, so $|k_0 - k_0'| \ll | \bm k - \bm k'|$.\footnote{While it may seem peculiar to integrate out massless states, the potential modes are off-shell. 
Moreover, the EFT contains ultra-soft modes with energy and momenta of order $|\bm k - \bm k'|$ but these encode dissipative effects irrelevant to the conservative potential.} For a classical system, the NR particles are separated by a distance $ |\bm r| \sim  1/|\bm k - \bm k'|$ that is parametrically larger than the Compton wavelengths of the particles, $ |\bm k|,|\bm k'|$.  The resulting hierarchy, $|\bm k - \bm k'| \ll |\bm k| , |\bm k'| $, corresponds to an expansion in large angular momentum, $J \sim | \bm  k \times \bm r|  \gg 1$, as utilized by Damour \cite{Damour:2016gwp,Damour:2017zjx} .    The classical component of any quantity is then extracted via the scaling
\eq{
  J^{-1} \propto \bm k - \bm k'\propto \kappa^{-1},
}{eq:hbar_scaling}
where ${\bm k}, {\bm k'} \propto 1+ J^{-1}$.  The first relation holds because angular momentum scales linearly with distance while the second relation holds due to the virial theorem.  Here $\kappa$ is the coupling constant, which for example in gravity is the gravitational constant, $\kappa = 4\pi G$.  The classical potential has the same scaling as the leading Coulomb interaction, $ \kappa / |\bm k - \bm k'|^2 \propto J^{3}$. 

Higher order potential terms are parametrized by arbitrary Hermitian combinations of the rotational invariants $\bm k^2$, $\bm k'^2$, and $\bm k \cdot \bm k'$.  However, since $\bm k^2 - \bm k'^2$  vanishes on-shell,  it can be eliminated by a field redefinition. Similarly,  \Eq{eq:L_int} has no energy dependence since energy can also be traded for $\bm k^2$ and $\bm k'^2$ via the equations of motion.  We thus choose a field basis in which $V$ only depends on $\bm k^2 + \bm k'^2$ and $|\bm k - \bm k'|$, so~\cite{Neill:2013wsa}
\eq{
V(\bm{k},\bm{k}') &=  \frac{\kappa}{|\bm k - \bm k'|^2}  \left(  c_1+ c_2 \kappa |\bm k-\bm k'| + \cdots \right) ,
}{eq:Vansatz}
where we have only included terms which are classical and thus scale as $J^3$ in accordance with \Eq{eq:hbar_scaling}, and the ellipsis denotes terms higher order in $\kappa$.\footnote{Higher order classical terms odd in $\kappa$ include factors of $\log |\bm k-\bm k'|$.} 
 $c_i \! \left(\frac{\bm k^2 + \bm k'^2}{2}\right)$ are momentum-dependent functions characterizing contributions at $i$th order in the coupling constant and all orders in velocity.  Here we make the usual assumption that there is a convergent velocity expansion.
 
 \medskip

\noindent {\bf Amplitudes.}  
From \Eq{eq:L_kin} and \Eq{eq:L_int} it is straightforward to obtain the Feynman rules,
\eq{
\begin{tikzpicture}[baseline={([yshift=-1ex]current bounding box.center)},decoration={
    markings,
    mark=at position 0.55 with {\arrow{>}}}]
\node (a) at (0,0){};
\node (c) at (2,0){};
\draw[thick] (a.center) edge[postaction={decorate}] (c.center);
\node[anchor=south] at (1,0.05) {$(k_0, {\bm k})$};
 \end{tikzpicture} \ &=  \frac{i}{k_0-\sqrt{\bm k^2 + m_{A,B}^2}+i 0} \,,  \\
\begin{tikzpicture}[baseline={([yshift=-.5ex]current bounding box.center)},decoration={
    markings,
    mark=at position 0.55 with {\arrow{>}}}]
\node (a) at (-0.9,0.5){};
\node (b) at (0.9,-0.5){};
\node (c) at (-0.9,-0.5){};
\node (d) at (0.9,0.5){};
\draw[thick] (a.center) edge[postaction={decorate}] (0,0);
\draw[thick] (0,0) edge[postaction={decorate}] (b.center);
\draw[thick] (c.center) edge[postaction={decorate}] (0,0);
\draw[thick] (0,0) edge[postaction={decorate}] (d.center);
\node[anchor=south] at (-0.55,0.4) {\phantom{-}$\bm k$};
\node[anchor=south] at (0.5,0.4) {\phantom{-}$\bm k'$};
\node[anchor=north] at (0.5,-0.4) {-$\bm k'$};
\node[anchor=north] at (-0.55,-0.4) {-$\bm k$};
\vspace*{-1cm}
 \end{tikzpicture} \ &= -i V(\bm k,\bm k') \,,
}{}
where from here on the $+i0$ prescription will be implicit.

We are interested in the scattering amplitude for a process where $\bm p$ and $\bm p'$ are the incoming and outgoing three-momenta in the center of mass frame, and $E_A$ and $E_B$ are the energies of the incoming particles,
\eq{
E_{A,B} &= \sqrt{\bm p^2 + m_{A,B}^2} =\sqrt{\bm p'^2 + m_{A,B}^2} \, .
 }{eq:EAB}
We define the total energy and the reduced energy ratio,
\eq{
E = E_A +E_B  \quad \textrm{and} \quad
\xi = \frac{E_A E_B}{(E_A +E_B)^2} \, .
}{eq:Exi}
Note that $0\leq \xi \leq 1/4$ and moreover $\xi$ and $E$ are dependent variables since $E_A$ and $E_B$ are related through \Eq{eq:EAB}.  We also define the momentum transfer $\bm q = \bm p -\bm p' \propto J^{-1}$, with classical scaling dictated by \Eq{eq:hbar_scaling}.

The EFT amplitude can either be organized in terms of the $\kappa$ expansion or in terms of loop orders, so
\eq{
\MEFT =  \sum_{i=1}^\infty \MEFT^{(i)} = \sum_{L=0}^\infty \MEFT^{L\textrm{-loop}} ,
}{}
where $\MEFT^{(i)}$ is at $i$th order in $\kappa$ and arises from Feynman diagrams at $i-1$ loops and below.

Since pair creation of matter particles is kinematically forbidden in the NR limit, the amplitude at $L$ loops is comprised purely of iterated bubbles, so
\eq{
\MEFT^{L\textrm{-loop}}& =
\begin{tikzpicture}[baseline={([yshift=-.5ex]current bounding box.center)},scale=0.6,decoration={
    markings,
    mark=at position 0.55 with {\arrow{>}}}]
\node (i1) at (0-0.7, 0.7){};
\node (i2) at (0-0.7,-0.7){};
\node (a) at (0,0){};
\node (b) at (2,0){};
\node (c) at (4,0){};
\node (d) at (6,0){};
\node (f1) at (6+0.7, 0.7){};
\node (f2) at (6+0.7,-0.7){};
\draw[thick] (i1.center) edge[postaction={decorate}] (a.center);
\draw[thick] (i2.center) edge[postaction={decorate}] (a.center);
\draw[thick] (a.center) edge[bend left=70,looseness=1.2,postaction={decorate}] (b.center);
\draw[thick] (a.center) edge[bend right=70,looseness=1.2,postaction={decorate}] (b.center);
\draw[thick] (b.center) edge[bend left=70,looseness=1.2,postaction={decorate}] (c.center);
\draw[thick] (b.center) edge[bend right=70,looseness=1.2,postaction={decorate}] (c.center);
\draw[thick] (c.center) edge[bend left=70,looseness=1.2,postaction={decorate}] (d.center);
\draw[thick] (c.center) edge[bend right=70,looseness=1.2,postaction={decorate}] (d.center);
\draw[thick] (d.center) edge[postaction={decorate}] (f1.center);
\draw[thick] (d.center) edge[postaction={decorate}] (f2.center);
\fill [white] (2.65,1) rectangle (3.35,-1);
\node at (3.07,0) {\large $\cdots$};
\node[anchor=south] at (-0.4,0.7) {\phantom{-}$\bm p$};
\node[anchor=north] at (-0.4,-0.7) {-$\bm p$};
\node[anchor=south] at (-0.4+1.4,0.7) {\phantom{-}$\bm k_1$};
\node[anchor=north] at (-0.4+1.4,-0.7) {-$\bm k_1$};
\node[anchor=south] at (-0.4+5.5,0.7) {\phantom{-}$\bm k_L$};
\node[anchor=north] at (-0.4+5.5,-0.7) {-$\bm k_L$};
\node[anchor=south] at (-0.4+5.5+1.4,0.63) {\phantom{-}$\bm p'$};
\node[anchor=north] at (-0.4+5.5+1.4,-0.63) {-$\bm p'$};
\end{tikzpicture}.
}{eq:eftamp}
For convenience, we merge each pair of matter lines into an effective ``two-body propagator'',
\eq{
\Delta(\bm k) &= i \int \frac{dk_0}{2\pi} \frac{1}{k_0-\sqrt{\bm k^2 + m_A^2 } }\frac{1}{E -k_0-\sqrt{\bm k^2 + m_B^2}}\\
&= \frac{1}{E- \sqrt{\bm k^2 +m_A^2}  - \sqrt{\bm k^2 +m_B^2}  },
}{eq:Delta}
where the second line is obtained by closing the contour in $k_0$ either upwards or downwards in the complex plane.  The contribution at $L$ loops is then
\eq{
\MEFT^{L\textrm{-loop}} &= - \int_{{\bm k_1} \cdots {\bm k_L}}
V(\bm p ,\bm k_1) \Delta(\bm k_1) 
 \cdots  \Delta(\bm k_L) V(\bm k_L, \bm p') \\
&= - \int_{{\bm k_1} \cdots {\bm k_L}}
{{\cal N}_{\rm EFT}^{L\textrm{-loop}} \over X_{1}^2 X_{2}^2  \cdots X_{L+1}^2 Y_1 Y_2 \cdots Y_L},
}{eq:MEFT_loop}
where in the second line we have substituted the internal loop momenta $\bm k_n$ for equivalent variables,
\eq{
X_{n} = |\bm k_{n-1} - \bm k_n| \qquad {\rm and} \qquad Y_n= \bm k_n^2 - \bm p^2 \,,
}{eq:MEFTsimp}
describing the momentum transfer at each vertex and the off-shellness of each pair of matter propagators, respectively. Here $\bm k_0 = \bm p$  and $\bm k_{L+1} = \bm p'$ and ${\cal N}_{\rm EFT}^{L\textrm{-loop}}$ is a regular function of $X_{n}$ and $Y_n$ obtained from the Laurent expansion of the first line of \Eq{eq:MEFT_loop} in those variables. 

The variables in \Eq{eq:MEFTsimp} have several advantages. First, since $X_{n}\propto J^{-1}$, $Y_n \propto J^{-1}+J^{-2}$ and $\int_{\bm k} \propto J^{-3}$ in accordance with  \Eq{eq:hbar_scaling}, we can trivially extract the classical contribution by expanding ${\cal N}_{\rm EFT}^{L\textrm{-loop}}$ in the limit of small $Y_n$, keeping terms through order ${\cal O}(Y_n^{L})$ and all orders in velocity.
Second, \Eq{eq:MEFT_loop} manifests all singularities from matter particles as simple poles in $Y_n$.  These singularities correspond to lower order iterated contributions that are infrared divergent.  As we will see, since these iterations must exactly cancel against similar terms in the full theory, they play no role in the determination of the $c_i$. On the other hand, terms which are regular in $Y_n$ produce rational functions of the kinematic variables after integration and do affect $c_i$. 

\section{Full Theory}

\noindent {\bf Scalar Integral Decomposition.}  We now decompose all full theory amplitudes into a basis of scalar functions of the external four-momenta.  Here $p_1 = (E_A ,\bm p) $ and $p_2 = (E_B ,-\bm p)$ are the incoming four-momenta while $p_3 =  (E_A , \bm p')$ and $p_4 = (E_B , -\bm p')$ are outgoing. Like before, we decompose the full theory amplitude,
\eq{
M = \sum_{i=1}^\infty M^{(i)},
}{}
where $M^{(i)}$ is the contribution at $i$th order in $\kappa$, which arises purely at $i-1$ loops in the full theory. Here we define $M$ to have NR normalization, so it is proportional to the usual relativistic amplitude $\widetilde M = 4E_A E_B M$.
At tree level, the relativistic tree amplitude is
\eq{
\kappa^{-1} \widetilde M^{(1)} &=   d_{\topbot} \II_{\topbot}  \,, 
}{}
where $d_{\topbot}$ is a function of the external kinematics and the scalar tree function is defined as
\eq{
\II_{\topbot} = \frac{1}{(p_1 - p_3)^2}  = -\frac{1}{\bm q^2} \, .
}{} 
Similarly, the one-loop amplitude is
\eq{
i \kappa^{-2} \widetilde M^{(2)} =  d_{\Box} \II_{\Box} +d_{\bigtriangledown} \II_{\bigtriangledown} +d_{\bigtriangleup} \II_{\bigtriangleup} +\cdots ,
}{}
where the ellipsis denotes rational, bubble, and crossed-box contributions which do not contribute classically.  The scalar basis integrals are
\eq{
 \II_{\Box} &= \int_k \frac{1}{(p_1-k)^2}\frac{1}{(k-p_3)^2}  \frac{1}{k^2 -m_A^2}\frac{1}{(p_1 + p_2-k)^2 -m_B^2} \\
  \II_{\bigtriangledown,\bigtriangleup} &= \int_k  \frac{1}{(p_1-k)^2}\frac{1}{(k-p_3)^2}  \frac{1}{k^2 -m_{A,B}^2}\,  
   }{eq:ft_int}  
where  $\int_k= \int {d^4 k \over (2\pi)^4}$. The scalar coefficients $d_\Box$, $d_{\bigtriangledown}$, and $d_{\bigtriangleup}$ are rational functions of the external kinematics. 

 \medskip

\noindent {\bf Reduction to Three-Dimensional Integrals.} To compare the EFT and full theory amplitudes at the integrand level, we reduce the four-dimensional integrals in \Eq{eq:ft_int} to three-dimensional integrals expressed in terms of $X_n$ and $Y_n$. Our approach is similar to the method of regions~\cite{Beneke:1997zp,Smirnov:2002pj} except with an alternative prescription for contour integrals.  While the relativistic one-loop integrals in \Eq{eq:ft_int} have been computed previously, the procedure outlined here is formulated with the expressed purpose of scaling mechanically to higher loop orders.

In terms of the variables defined in \Eq{eq:MEFTsimp}, the triangle integral in \Eq{eq:ft_int} is
\eq{
  \II_{\bigtriangledown} &=  \int_k \, \frac{1}{k_0^2- X_1^2+i0}\frac{1}{k_0^2-X_2^2+i0}   \frac{1}{k_0^2 + 2 E_{A} k_0 -Y_1+i0}\, ,
   }{eq:I_tri3d}
where we parameterize $k = (E_A + k_0, \bm k_1)$ so that $k_0$ describes deviations from an instantaneous potential. The classical potential is generated by off-shell mediators in the potential region with $ | k_0|  \ll X_{1,2} $. Thus, we consider the contribution to the integral in \Eq{eq:I_tri3d} from a contour on the real $k_0$ axis along this interval. 

We can evaluate this by pushing the contour either upwards or downwards, provided one includes non-zero contributions from the upper or lower arc.  Including relative signs from contour orientation, these arc contributions are equal and opposite and their difference is the residue at infinity, which is in general non-zero.  Thus, an equivalent but more convenient prescription is to take the average result from pushing the contour upwards and downwards, {\it i.e.}~half the sum of all residues enclosed by the full circle, including signs from orientation.  Crucially, this region does {\it not} contain any poles from mediator propagators, since $|k_0| \ll X_{1,2}$.  Furthermore, while the matter propagator contains both a particle and anti-particle pole, at most one can lie in the potential region.  So {\it e.g.}~in \Eq{eq:I_tri3d} we would only include the pole at $k_0 = \sqrt{E_A^2 +Y_1} - E_A - i0$ because it vanishes in the instantaneous limit, $Y_1=0$.
The upshot is that the energy integral from the potential region effectively yields
\eq{
\int \frac{dk_0}{2\pi} (\cdot) ={i \over 2}  \left[  \sum_{k_* \in \mathbb{H^+}} \!\!\underset{\,\, k_0 = k_* }{{\rm Res}}(\cdot)-  \sum_{k_* \in \mathbb{H^-}} \!\!\underset{\,\, k_0  = k_*}{{\rm Res}}(\cdot) \right],
}{}
where the sum runs over residues $k_*$ from potential region matter poles in the upper/lower half planes, $\mathbb{H}^\pm$. This prescription is equivalent to applying key identities used in showing the exponentiation of the eikonal amplitude~\cite{Saotome:2012vy,Akhoury:2013yua}.

After performing the $k_0$ integral we expand the remaining three-dimensional integrand in the NR limit of large $m_{A,B}$.  We then extract the classical contribution according to the $J$ power counting discussed below~\Eq{eq:MEFTsimp}. 
For example, the triangle integrals in~\Eq{eq:ft_int} depend only on the four-momenta $p_1$ and $p_1 - p_3$, whose on-shell inner products are functions only of $|\bm q|$ and $m_{A,B}$.  Hence, the NR expansion is a power series in $|{\bm q}|/m_{A,B}$ and since $\bm q \propto J^{-1}$ it is obvious that the classical term coincides with the leading term in the large mass expansion. Similarly, for the box integral we expand in large $E_{A,B}$.

In summary, the scalar integrals can be written as
 \eq{
 \II_{\Box} &= \frac{i}{2E} \int_{\bm k}
  \,  
  {1 \over X_1^2 X_2^2 Y_1} + \cdots \,, \\
  \II_{\bigtriangledown,\bigtriangleup} &= -\frac{i}{4  m_{A,B}} 
  \int_{\bm k}
   \,   
   {1 \over X_1^2 X_2^2} + \cdots   \,, 
}{eq:ft_int3d}
where the ellipses denote contributions which are higher order in $J^{-1}$ and thus quantum, and $\int_{\bm k} {1 \over X_1^2 X_2^2} = {1 \over 8 |\bm q|}$ by standard integral formulas~\cite{Smirnov:2004ym}. 
Including the coupling constant, we find that $\kappa^2  \II_{\Box}\propto J^4 $ and  $\kappa^2    \II_{\bigtriangledown,\bigtriangleup} \propto J^3 $, so the triangle is classical but the box is actually superclassical since it encodes iterations of the tree-level potential that will cancel with similar terms in the EFT.   

\section{Matching Calculation}\label{sec:match}
The potential coefficients $c_i$ are obtained by matching the EFT and full theory amplitudes order-by-order in $\kappa$,  so $ M^{(i)} - \MEFT^{(i)}=0$. This procedure is greatly simplified by expressing this difference of amplitudes at the integrand level, since terms with poles in $Y_n$ which  evaluate to infrared non-analyticities are canceled without performing complicated integrals. 
This cancelation occurs because the EFT and full theory have identical cut structure at low energies, as mandated by the starting assumption that the theories describe the same infrared dynamics.   This holds at all loops, provided all relevant momentum regions have been included in the EFT.

That such a subtraction can be done at the integrand level should not be obvious because loop momenta in distinct diagrams generally have ambiguous relative orientation since there is no intrinsic origin in loop momentum space. Crucially, in our case the integrands can be aligned by matching their $Y_n$ poles. The remaining terms then trivially integrate to rational functions of the external kinematics. 

At leading and next-to-leading order in $\kappa$ we find
\eq{
 M^{(1)}- \MEFT^{(1)} &= \frac{\kappa}{\bm q^2}  \left[c_1(\bm p^2)-  \frac{d_{\topbot}}{4E^2 \xi }  \right]\,, \\
 M^{(2)}- \MEFT^{(2)}  &= \kappa^2  \left[ \frac{c_2(\bm p^2)}{|\bm q|}  
 + 
 \int_{ \bm k} 
  {  {\cal N}^{1\textrm{-loop}} - {\cal N}_{\rm EFT}^{1\textrm{-loop}} \over X_1^2 X_2^2 Y_1} \right] \, ,
}{eq:2PM_match}
where the EFT and full theory integrand numerators are
\eq{
{\cal N}_{\rm EFT}^{1\textrm{-loop}} &=   \left[  2 E\xi   +  Y_1\left(\frac{1-3\xi}{2 E \xi} +  E \xi \partial_{\bm p^2}  \right) \right] c^2_1(\bm p^2) \,,  \\
{\cal N}^{1\textrm{-loop}} &= \frac{d_{\Box}}{8E^3 \xi} - \frac{Y_1}{16E^2 \xi}\left( \frac{d_{\bigtriangledown}}{m_A}+ \frac{d_{\bigtriangleup}}{m_B} \right) \, .
}{}
Since the left-hand sides of \Eq{eq:2PM_match} are zero, we can solve explicitly for $c_1$ and $c_2$. We find the following solutions, which apply to all orders in velocity:
\begin{widetext}
\eq{
c_1(\bm p^2) =  \frac{d_{\topbot}}{4E^2 \xi} \qquad {\rm and} \qquad
c_2(\bm p^2) = \frac{1}{128 E^2 \xi} \left[ \frac{\left( - 1+ \xi +2E^2\xi^2 \partial_{\bm p^2} \right) d_{\topbot}^2}{2E^3 \xi^2}    + \frac{d_{\bigtriangledown}}{m_A}+ \frac{d_{\bigtriangleup}}{m_B} \right] \,. 
}{eq:result_2PM}
\end{widetext}
Note that $c_2$ is simply a rational function of $\bm p^2$ since, as discussed above, terms in the integral that have poles in $Y_1$ 
that would yield infrared logarithms
cancel exactly at the integrand level.
In particular, the ${\cal O}(Y_1^0)$ term in the difference of numerators is
\eq{
{\cal N}^{1\textrm{-loop}}-{\cal N}_{\rm EFT}^{1\textrm{-loop}}  &= \frac{1}{8E^3  \xi}\left[  d_\Box-d^2_{\topbot}  \right] + {\cal O}(Y_1) \, ,
}{eq:N_diff}
which implies that $ d_\Box = d^2_{\topbot}$. Indeed, this relation is obvious from the point of view of unitarity in the full theory, since the coefficient of the scalar box integral is given by the product of tree amplitudes.

\section{Gravity Results}

We have computed the classical potential at leading and next-to-leading order, $c_1$ and $c_2$ in \Eq{eq:result_2PM}, expressed in terms of the scalar functions $d_{\topbot}$, $d_{\bigtriangledown}$, and $d_{\bigtriangleup}$ that are the natural outputs of a generalized unitarity calculation.  For GR, both the full amplitude calculation~\cite{BjerrumBohr:2002kt} as well as the unitarity calculation have been completed~\cite{Neill:2013wsa,Cachazo:2017jef, Guevara:2017csg,Bjerrum-Bohr:2018xdl}, yielding
\eq{
d_{\topbot} &= 4 \left[ m_A^2 m_B^2 - 2 (p_1 \cdot p_2)^2  \right] \,, \\
d_{\bigtriangledown,\bigtriangleup} &= 12m_{A,B}^2 \left[ m_A^2 m_B^2 -5 (p_1 \cdot p_2)^2 \right] \,,
}{eq:dGR}
where $p_1 \cdot p_2 =   E_A E_B+\bm p^2$ and we have kept only classical contributions. Note that these quantities can also be constructed with the aid of color-kinematics duality~\cite{Bern:2008qj,Bern:2010yg,Bern:2010ue}, provided one can eliminate unphysical dilaton and axion modes (see also Refs.~\cite{Monteiro:2014cda,Monteiro:2015bna,Luna:2016hge,Goldberger:2016iau,Ridgway:2015fdl,Luna:2016due}  for the related classical double copy).
Inserting these into \Eq{eq:result_2PM} yields
\eq{
c_1(\bm p^2) &= {1 \over E^2 \xi} \left[ m_A^2 m_B^2 - 2 (p_1 \cdot p_2)^2  \right] \,,  \\
c_2(\bm p^2) &= {1 \over 32 E^2 \xi} \left[ 2 E (\xi-1) c^2_1(\bm p^2)  -16 E (p_1 \cdot p_2) c_1(\bm p^2) \right. \\
&\quad \left. + 3 (m_A +m_B)  (m_A^2 m_B^2 -5 (p_1 \cdot p_2)^2 )   \right] \,,
}{eq:result_gravity_2PM}
which specify the potential at 2PM, {\it i.e.}~${\cal O}(G^2)$ and to all orders in velocity. In position space, our classical potential in the center of mass frame is 
\eq{
V(\bm p, \bm r) &=\frac{G c_1(\bm p^2)}{|\bm r|}+ \frac{8G^2 c_2(\bm p^2)}{\bm r^2} + \cdots,
}{eq:Vpos}
where $\bm r$ is the distance vector between the black holes. It would be interesting to connect our results to those of Damour~\cite{Damour:2016gwp,Damour:2017zjx}, who has used effective one-body methods to compute the 2PM potential from scattering angles (for calculations of the 1PM potential see Refs.~\cite{Ledvinka:2008tk,Foffa:2013gja}).

Since these expressions are valid at all orders in velocity they can be compared against state of the art calculations that extend up to 4PN order.  A nontrivial complication is that our potential in \Eq{eq:Vpos} differs from those in Refs.~\cite{Bernard:2015njp,Jaranowski:2015lha} by a gauge transformation.  While this gauge transformation can in principle be constructed \cite{Holstein:2008sx}, this requires considerable effort. Here we employ a new approach: we instead compute the on-shell scattering amplitude $\MEFT$ for the two potentials under comparison.  Since the amplitude encodes all the relevant dynamics and is gauge invariant, they will match provided the potentials are gauge equivalent.  Calculating $\MEFT$ for our potential and comparing it to $\MEFT$ computed from the known potentials given in Eq.~(223) of \Ref{Blanchet:2013haa} and Eq.~(8.41) of \Ref{Jaranowski:2015lha}, we obtain exact agreement including all terms through ${\cal O}(G^2)$. We have also checked that in the limit $m_A/m_B \gg 1$ our result agrees with the potential for a test body orbiting a Schwarzchild black hole to ${\cal O}(G^2)$ and all orders in velocity~\cite{Jaranowski:1997ky}.

Given that the on-shell amplitude is unique and the classical potential is not, it may seem strange that our construction extracts a unique expression for the latter from the former. However, recall that our starting point in \Eq{eq:Vansatz} does not include terms that can be eliminated by equations of motion, which in itself is the choice of gauge~\cite{Manohar:2000hj}. Such terms do not affect on-shell amplitudes, but enter in loops and change the resulting potential coefficients in a way that is a pure gauge transformation.

\section{Conclusions}

We have derived a systematic map between the classical potential for spinless particles and their corresponding on-shell scattering amplitudes.   
Our main result, summarized in \Eq{eq:result_2PM}, relates the classical potential coefficients $c$, describing effects at leading and next-to-leading order in the coupling constant and all orders in velocity, to the scalar coefficients $d$, which are the natural output of a unitarity calculation of the on-shell scattering amplitude in a general QFT. For the special case of GR, we have verified agreement of our results, summarized in \Eq{eq:result_gravity_2PM} and \Eq{eq:Vpos}, to all known formulas in the literature, which extend up to 4PN order.  

Our current results may have utility for future work on gravitational waves. Since \Eq{eq:result_2PM} includes all orders in velocity, it will serve as a check of higher order PN calculations. Furthermore, it is applicable to any QFT, so it allows for modifications of GR involving new light states or higher dimension operators.

This paper leaves several promising avenues for future work.  Foremost is the extension to next-to-next-to-leading order in the coupling constant, which for gravity is 3PM order. While our methods are by construction scalable to higher orders, new subtleties may emerge, {\it e.g.}, from the logarithms of momentum in the 3PM potential.  Moreover, we will eventually encounter infrared divergences due to overlap from radiation modes at 4PN~\cite{Porto:2017dgs} as well as ultraviolet divergences which formally enter at 3PN and affect physical observables at 5PN when finite size effects become relevant.  Application of our methods to particles with spin and for gravitational wave emission should also be interesting.

{\it Acknowledgments:}
We thank Zvi Bern and Chia-Hsien Shen for comments on the manuscript and useful discussions. C.C.~is supported by a Sloan Research Fellowship and DOE Early Career Award under grant no. DE-SC0010255. I.Z.R.~is supported by the NSF under grant no. NSF-1407744.  M.P.S.~is supported by the DOE under grant no. DE-SC0011632 and the McCone Fellowship at the Walter Burke Institute.

\end{document}